\documentclass[aps,twocolumn,showpacs,preprintnumbers,amsmath,amssymb,nofootinbib,superscriptaddress,showkeys]{revtex4-1}
\usepackage{amsmath,epsfig}
\usepackage{longtable}
\allowdisplaybreaks[4] 

\usepackage[normalem]{ulem}
\usepackage{color}

\newcommand{\bear}{\begin{eqnarray}}
\newcommand{\eear}{\end{eqnarray}}

\newcommand{\ba}{\begin{array}{c}}
\newcommand{\ea}{\end{array}}

\newcommand{\be}{\begin{equation}}
\newcommand{\ee}{\end{equation}}

\begin{document}



\preprint{MITP/14-070}
\preprint{IFT-UAM/CSIC-14-100}
\preprint{FTUAM-14-38}

\title{Precise determination of resonance pole parameters through Pad\'e approximants}

\author{Pere Masjuan} \email{masjuan@kph.uni-mainz.de}
 \affiliation{Institut f\"ur Kernphysik, Johannes Gutenberg-Universit\"at, D-55099 Mainz,
 Germany }

\author{Jacobo Ruiz de Elvira} \email{elvira@hiskp.uni-bonn.de}
\affiliation{Helmholtz-Institut f\"ur Strahlen- und Kernphysik, Universit\"at Bonn, D-53115
Bonn, Germany}

\author{Juan Jos\'e Sanz-Cillero} \email{juanj.sanz@uam.es}
\affiliation{Departamento de F\'isica Te\'orica and Instituto de F\'isica Te\'orica, IFT-UAM/CSIC
       Universidad Aut\'onoma de Madrid, Cantoblanco, Madrid, Spain}


\begin{abstract}
In this work, we present a precise and model--independent method to extract resonance pole parameters from phase-shift scattering data.
These parameters are defined from the associated poles in the second Riemann sheet, unfolded by the analytic continuation to the complex pole
using Pad\'e approximants. Precise theoretical parameterizations of pion-pion scattering phase shifts based on once-- and twice-- subtracted dispersion relations are used as input, whose functional form allows us to show the benefit and accuracy of the method.
In particular, we extract from these parametrization the pole positions of the
  $f_0(500)$ at $\sqrt{s}=(453\pm 15) - i(297 \pm 15)$~MeV, the $\rho(770)$ at $\sqrt{s}=(761.4\pm 1.2)-i(71.8 \pm 1.0)$~MeV, and the pole of the
  $f_2(1270)$, located at $\sqrt{s}=(1267.3\pm 1.7)-i(95.0\pm 2.3)$~MeV.  
The couplings of the resonances to two pions are also determined with high
precision, obtaining respectively, $3.8\pm 0.4$~GeV, $5.92\pm 0.15$ and
$4.41\pm 0.23$~GeV$^{-1}$.
Special attention is dedicated to the systematic treatment of the theoretical and statistical uncertainties, 
together with their comparison with previous determinations.

\end{abstract}

\pacs{11.55.-m,11.80.Fv,12.40.Vv,12.40.Yx,13.40.Gp,14.40.-n}

\keywords{Pad\'e Approximants, Resonance poles and properties}

\maketitle


\setcounter{footnote}{0}

\section{Introduction}

The non-perturbative regime of Quantum Chromodynamics is characterized by the presence of hadronic resonances
defined by complex $S$--matrix poles in unphysical Riemann sheets.
Contrary to other definitions, the pole position --and the corresponding
pole mass and width defined by $s_p=(M_p-i\Gamma_p/2)^2$-- is universal
and independent of the process under consideration. In addition, its residue enclose the information on the underlying process.

However, extrapolating the physical amplitude at real values of the energy,
i.e., in the 1st Riemann sheet, into the complex plane
and extracting resonance poles is not a trivial task.
The extrapolation procedure may change drastically the value of the outcomes, specially
in the case of broad states.

The simple method proposed here for the analytical continuation
is given by the Pad\'e approximants (PA) to an amplitude $F(s)$
in terms of the total invariant squared momentum $s$ around a point $s_0$, denoted by $P^N_M(s,s_0)$~\cite{Baker}:
\begin{equation}
\label{PAdef}
P^N_M(s,s_0)\,=\,  F(s)\,\,\, +\,\,\, {\cal O}\bigg((s-s_0)^{M+N+1}\bigg)\,,
\end{equation}
with $P^N_M(s,s_0)=Q_N(s)/R_M(s)$ given by the ratio of two polynomials $Q_N(s)$
and $R_M(s)$ of degrees $N$ and $M$, respectively~\cite{Baker}.
$R_N(s_0)$ is chosen to be $1$,  without any loss of generality.

A special case of interest for the present work is given by
Montessus de Ballore's theorem~\cite{Montessus,Masjuan:2013jha}.
Its simpler version states that when the amplitude $F(s)$ is analytic
inside the disk $B_\delta(s_0)$ except for a single pole at $s=s_p$ the sequence  of one-pole PA $P_1^N(s,s_0)$,
\begin{equation}\label{PAeq}
P_1^N(s,s_0)=\sum_{k=0}^{N-1}a_k(s-s_0)^k+\frac{a_N(s-s_0)^N}{1-\frac{a_{N+1} }{a_N} (s-s_0)}\, ,
\end{equation}
converges to $F(s)$ in any compact subset of the disk excluding the pole $s_p$.
%
The constants $a_n=\frac{1}{n!}F^{(n)}(s_0)$
are given, accordingly, by the $n^{th}$~derivative of
$F(s)$~\cite{Montessus,Baker,Masjuan:2013jha}, being $P^N_1(s,s_0)$ determined by the first
derivatives $F^{(0)}(s_0)=F(s_0)$, $F^{(1)}(s_0)$... $F^{(N+1)}(s_0)$.

Likewise, the PA pole  and residue
\bear
s_p^{(N)}=s_0+\frac{a_N}{a_{N+1}}\, , \,\,
Z^{(N)}=\, - \frac{(a_N)^{N+2}}{(a_{N+1})^{N+1}}\, ,
\label{eq:PN1-predictions}
\eear
converge to the corresponding pole and residue of $F(s)$ for $N\rightarrow \infty$.

During the last years, dispersive approaches have been proved to be a very successful
tool to obtain precise determinations of phase shifts and pole
parameters~\cite{Colangelo:2001df,Moussallam:2004,Caprini:2005zr,DescotesGenon:2006uk,Martin:2011cn,GarciaMartin:2011jx,Ditsche:2012fv}.
However, they are based on a complicated although powerful machinery which makes
them difficult to use except for a limited number of cases.
In this letter, we use dispersive $\pi\pi$ parameterizations to show how it is possible to obtain a precise and model-independent determination of resonance pole parameters using the theory of PA~\cite{Baker,Masjuan:2013jha}, even for cases where dispersive methods cannot be easily applied.

Following the proposal in Ref.~\cite{Masjuan:2013jha}, Montesus' theorem is applied to the simplest case with a single-resonance
pole inside the disk $B_{\delta}(s_0)$. Nonetheless, it can be generalized, ensuring the convergence of the $P^N_M(s,s_0)$ sequence
when the amplitude contains up to $M$ poles in the disk $B_{\delta}(s_0)$~\cite{Masjuan:2013jha}.

Thanks to  Montessus' theorem, one can  use the PAs in a theoretically safe way
by centering them at $s_0+i0^+$ over a physical brunch cut
and far away enough from the branch point singularities,
which will limit the theorem's applicability range in the $s$--variable.
This allows us to unfold the 2RS, or higher sheets,  through  the analytical extension of $F(s)$ from
the first Riemann sheet (1RS) provided by the PA~\cite{Masjuan:2013jha}.

As the order of the approximant increases,  the difference between consecutive orders become
smaller, and the $s_p^{(N)}$ predictions defined in Eq.~\eqref{eq:PN1-predictions} converge to  the actual pole $s_p$ of
the amplitude $F(s)$.
Therefore, we will consider the difference between the $P^N_1(s,s_0)$ and
$P^{N-1}_1(s,s_0)$  as our estimator of the systematic theoretical error for $s_{p}^{(N)}$~\cite{Masjuan:2013jha}:
\bear
\label{eq:error}
\Delta s_N  \equiv   | s_{p}^{(N)}  -  s_{p}^{(N-1)}| ,
\,\,
\Delta Z_N  \equiv   | Z^{(N)}  -  Z^{(N-1)}| .
\eear
Several examples in phenomenological models together with rates of convergence for Eq.~\eqref{eq:error}
can be found in Ref.~\cite{Masjuan:2013jha}.

\section{$\pi\pi$-scattering and poles}\label{sec:pipi-data}

The success of our pole position determinations will rely on our capability
to obtain a precise determination of the
coefficients $a_j$ appearing in Eq.~\eqref{eq:PN1-predictions},
i.e., a sequence of $nth$--order derivatives with respect
to $s$ for the partial-wave at a given point.

\begin{figure}
  \includegraphics[width=3.25in]{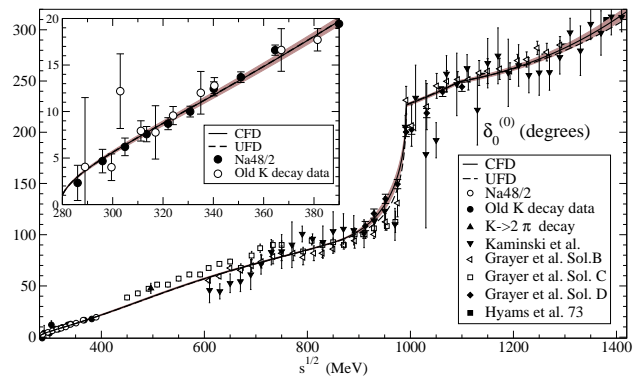}
  \caption{S0 wave phase shift for $\pi\pi$--scattering
  experimental data together with the UFD and CFD parameterizations~\cite{Martin:2011cn}.
  The dark band covers the uncertainties. In the inner top panel,
  we show the low-energy region and the good description of
  the latest NA48/2 data on $K_{l4}$ decays,
  which are responsible for the small uncertainties of the UFD and CFD parameterizations.}
  \label{fig:S0wave}
\end{figure}

In this work we use the recent and very precise output of the $\pi\pi$
scattering data analysis performed in~\cite{Martin:2011cn}.
This analysis incorporates $\pi\pi$ scattering and $K_{l4}$ decay data
--in particular, the latest results from NA48/2 \cite{Batley:2010zza}--,
obtaining, as a first step, a simple set of unconstrained parametrization (UFD)
fitted to these data for each partial wave separately up to 1.42~GeV.
Consequently, this UFD parametrization is used as a starting point
for a Constrained Fit to Data (CFD),
in which forward dispersion relations, Roy equations,
and one-subtracted coupled partial wave dispersion relations --or GKPY equations--
are imposed as an additional constraint to the data fits.
These relations incorporate crossing and assume analyticity in the 1RS.
The interest of these CFD parameterizations is that, while describing the data,
they satisfy within uncertainties dispersion relations,
constraining and  reducing the errors of the experimental input.
This is shown in Fig.~\ref{fig:S0wave},
where the resulting scalar-isoscalar $\pi\pi$ phase-shift is presented.
Both the UFD and CFD describe the experimental data, but in addition,
the CFD satisfies the dispersive constraints imposed.

The high accuracy obtained in this dispersive analysis gives us the opportunity to use the CFD parametrization as input to obtain a precise determination
of the coefficients $a_j$ in Eq.~\eqref{eq:PN1-predictions}, and then, to extract the pole position of the lightest resonances appearing in $\pi\pi$-scattering in the $IJ=00,11,02$ channels, respectively,   i.e., the $f_0(500)$, the $\rho(770)$, and the $f_2(1270)$.
Furthermore, these parameterizations were used in~\cite{GarciaMartin:2011jx}
as input for the GKPY and Roy S0- and P-wave equation
for $\pi\pi$--scattering,
providing a model-independent continuation to the complex plane, and then, a determination
of the position and residues of the second Riemann sheets poles appearing in these channels,
which we can use to compare the precision of our pole extraction method, and the analysis of the errors.

\begin{figure}
 \includegraphics[width=3.25in]{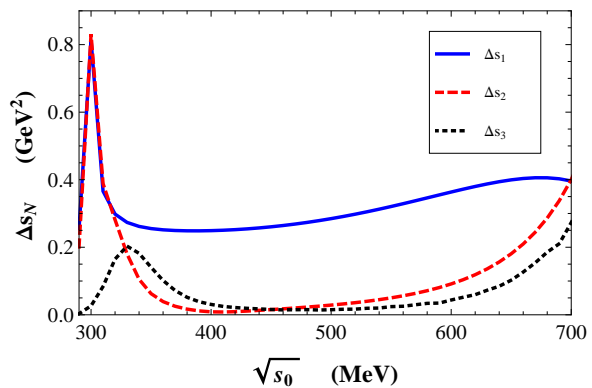}
\caption{{\small
Uncertainty $\Delta s_{N}$ in the $\sigma$ pole determination
for the $P^1_1(s,s_0)$ (solid blue), $P^2_1(s,s_0)$ (dashed red)
 and $P^3_1(s,s_0)$ (dotted black) approximants for different values of $s_0$ ranging from $2 m_{\pi}$ up to $700$ MeV. The fastest convergence is found around $500$ MeV.
  }}
 \label{fig.convergence}
\end{figure}
\begin{figure}
  \includegraphics[width=2.25in]{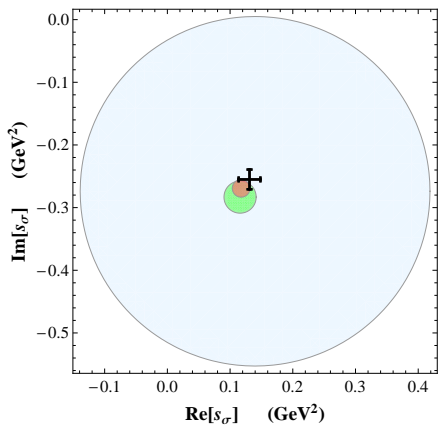}
  \caption{{\small
 Theoretical uncertainty regions $\Delta s_{N}$ for
 the $P^1_1(s,s_0)$ (lighter blue), $P^2_1(s,s_0)$ (green)
 and $P^3_1(s,s_0)$ (darker red) approximants.  The  $1\sigma$ black error bar corresponds to the
 determination through the GKPY equations~\cite{GarciaMartin:2011jx}.
 The PA center in this plot is $\sqrt{s_0}=490$~MeV.
  }}
  \label{fig.theory-CL-regions-sigma}
\end{figure}

Now, let us be more precise with our method and proceed to the analysis of resonances in various channels, beginning with the $f_0(500)$ or $\sigma$ meson.

We use the CFD $\pi\pi$ parameterizations to obtain the value of the phase-shift $\delta^0_0(s)$ and inelasticity $\eta^0_0(s)$,
as well as their four first derivatives. From them, we compute the value and derivatives of the $IJ=00$ partial wave,
\begin{equation}
t^I_J(s)=(\eta^I_J(s) e^{2i\delta^I_J(s)}-1)/(2i\rho_\pi(s)),
\end{equation}
where  $\rho_{\pi}(s) = \sqrt{1-4m_{\pi}^2/s}$ is the phase-space factor,

Our method, then, proceeds as such.
First, from the central values of the $\delta^0_0(s)$ $\pi\pi$ phase-shift (Fig.~\ref{fig:S0wave}) and its derivatives,
we  analyze the convergence of the theoretical uncertainty
$\Delta s_N$ of the $P^N_1(s,s_0)$ approximants $N=1,2,3$ for different PA centers $s_0$ between the $\pi\pi$ and $KK$ thresholds.
Note that below the $KK$ threshold, elastic scattering is assumed  for the S0 wave in~\cite{GarciaMartin:2011jx}, so we take $\eta^0_0(s)=1$.

Afterwards, the theoretical error $\Delta s_3$ for the $P^3_1(s,s_0)$ approximant
happens to be minimized at $\sqrt{s_0}=490$~MeV, see Fig.~\ref{fig.convergence}, and the PA sequence is found to break down
when $s_0$ approaches either the $\pi\pi$ or $KK$ thresholds.
In this way, we are able to obtain a first estimate from the $P^3_1(s,490^2 \textrm{MeV}^2)$ without including the uncertainties of the CFD parametrization:
\bear
\sqrt{s_\sigma} =  (453\,\pm \, 13_{\textrm{sys}})+i(297 \pm 13_{\textrm{sys}}) \mbox{~MeV} \,.
\eear
Even for such a broad resonance, a clear convergence
can be observed in Fig.~\ref{fig.theory-CL-regions-sigma},
where we plot the $s_p$ theoretical uncertainty regions for the different $P^N_1(s,s_0)$.

Finally, in order to incorporate the statistical uncertainties coming from the input error bands,
we use a MonteCarlo (MC) simulation, where for each $s_0$ between the $\pi\pi$ and $KK$ thresholds, $\{ \delta^{(n)}  \} $  configurations (with $n=0$--$4$)
are generated with a distribution according to the values of the phase-shift
and derivatives of~\cite{Martin:2011cn}.
However, the theoretical error in $P^3_1(s,s_0)$ is not negligible anymore and each of these MC configurations for $\{ \delta^{(n)}  \} $  do not correspond to a single point
$s_p^{(3)}$,  but to an homogeneous circle centered at that point with radius
$\Delta s_3$, as in Fig.~\ref{fig.theory-CL-regions-sigma}.
Practically, from every point $s_p^{(3)}$ produced, the MC generates a fixed number $n$ of points with a uniform distribution within the circle of radius $\Delta s_3=|s_p^{(3)}-s_p^{(2)}|$, centered at the given  $s_p^{(3)}$.

The theoretical error due to the truncation of the PA sequence is
of the same order as the uncertainties that stem just from the error of the
phase-shift data of~\cite{Martin:2011cn}.

\begin{table}[!h]   
\caption{Collection of different dispersive predictions for the $f_0(500)$ meson pole
  position and coupling to two pions in the 2RS. }
\begin{center}
\begin{tabular}{ccc}
\hline
Reference& $\sqrt{s_\sigma}$(MeV)&$|g_{\sigma\pi\pi}|$ (GeV)\\
\hline
 \cite{Colangelo:2001df}  
&     $(470\pm 30) -  i (295 \pm 20)$   
&   --      \\
   \cite{rho-Zheng}    
&   $(470 \pm 50)-i(285 \pm 25)$
&  --    \\
\cite{Caprini:2005zr}
&   $(457^{+14}_{-13}) -  i(272^{+9}_{-12.5})$&$3.31^{+0.35}_{-0.15}$\\
\cite{GarciaMartin:2011jx}
&  $(457^{+14}_{-13})  -  i(279^{+11}_{-7})$ & $3.59^{+0.11}_{-0.13}$\\
\cite{Moussallam:2011zg}
&  $(442^{+5}_{-8}) -  i(274^{+6}_{-5})$&$3.37$\\
\hline
\mbox{This Work} 
&  $(453\pm      15         ) -   i(297 \pm 15)$
&   $3.8\pm0.4$
\\
\hline
\end{tabular}
\end{center}
\label{tab:sigma-pole}
\end{table}

The combined error (theory+experiment) from the MC is minimal for $\sqrt{s_0}=500$~MeV
and the $P^3_1(s,s_0)$ approximant produces the $f_0(500)$ pole position shown in Table~\ref{tab:sigma-pole}.
In addition, we also provide in Table~\ref{tab:sigma-pole} the $\sigma$ coupling to two pions defined as
\begin{equation}
g^2=-16\pi Z^{(N)}(2l+1)/(2p)^{2l},
\end{equation}
where $p^2=s/4-m_\pi^2$,  and $Z^{(N)}$ is the pole residue given in
Eq.~\eqref{eq:PN1-predictions}, which we calculate from the $P^3_1(s,s_0)$ in a similar way ($|Z^{(3)}|=0.30\pm 0.06$~GeV$^2$).
We also show in Table~\ref{tab:sigma-pole} further $f_0(500)$ determinations.
In particular, the comparison with the result of~\cite{GarciaMartin:2011jx} is specially illuminating,
since it was obtained from the analytic continuation to the complex plane of the GKPY eqs,
 using as input, the same CFD parameterizations employed in this work.
The agreement between both determinations highlight the goodness of PA
as a precise method to extract resonance pole parameters.

In order to analyze the $\rho(770)$ resonance, we repeat exactly the same procedure
for the $I=J=1$ channel, using, this time, the CFD $\delta^1_1(s)$ parametrization of~\cite{Martin:2011cn}.
The inelasticity is again taken as $\eta=1$ below the $KK$ threshold.
Without input errors, the optimal point for the PA center is
$\sqrt{s}_0=680$~MeV, where one finds a fast convergence
for the $P^N_1(s,s_0)$ sequence:
$\Delta s_1= 4.7\cdot 10^4$~MeV$^2$, $\Delta s_2=1.0\cdot 10^3$~MeV$^2$,
$\Delta s_3=4.1$~MeV$^2$.  Furthermore,
the $\rho$ pole position uncertainties are below $0.1$~MeV for $P^3_1(s,s_0)$
if $\sqrt{s_0}\in [\, 0.65$~GeV$\, ,\, 0.8$~GeV$\, ]$.
Incorporating the uncertainties of the parametrization through a MC, as we did before
for the $f_0(500)$,  we find that the combined error (theory+experiment)
is minimized at $\sqrt{s}_0=740$~MeV giving the $\rho$ pole position of Table~\ref{tab:rho-pole}.
Similar outcomes  are  obtained for the range $730$--$780$~MeV up to $0.2$~MeV variations
in the error size.
We also show in Table~\ref{tab:rho-pole} the $\rho$ coupling to two pions,
extracted again from the pole residue ($|Z^{(3)}|=0.118\pm 0.006$~GeV$^2$).
Contrary to what happened with the $f_0(500)$, in the case of the $\rho(770)$  most
of the error comes from the input uncertainties being the theoretical error essentially negligible.
As in the case of the $f_0(500)$, the comparison with~\cite{GarciaMartin:2011jx} shows
a perfect agreement.
\begin{table}[!t]
\caption{{\small Collection of different dispersive predictions for the $\rho$ meson
resonance parameters  in the 2RS.}}
\begin{center}
\begin{tabular}{ccc}
\hline
Reference& $\sqrt{s_{\rho}}$ (MeV) & $|g_{\rho\pi\pi}|$\\
\hline
\cite{Roy} &  $(762.5\pm 2)- i(71 \pm 4)$
& -- \\
 \cite{Colangelo:2001df} 
&   $(762.4 \pm 1.8) - i(72.6 \pm 1.4)$
&  --   \\
 \cite{Cillero-VFF} &   $(764.1\pm 2.7^{+4.0}_{-2.5})-i(74.1\pm 1.0^{+0.9}_{-3.0}$ 
&  --   \\
 \cite{rho-Zheng}
&   $(763.0\pm 0.2)-i(69.5\pm 0.3)$ 
&    --    \\
\cite{GarciaMartin:2011jx}
&  $(763.7^{+1.7}_{-1.5}) -i(73.2^{+1.0}_{-1.1})$
& $6.01^{0.04}_{-0.07}$\\
  \cite{Masjuan:2013jha} 
&   $(763.7\pm 1.2) - i (   72.0 \pm1.5)$
&      --     \\
\hline
\mbox{This Work} & $(761.4\pm 1.2)-i(  71.8  \pm 1.0)$
&  $5.92\pm 0.15 $\\
\hline
\end{tabular}
\end{center}
\label{tab:rho-pole}
\end{table}

Finally, we want to end up with the study of the isoscalar
tensor resonance $f_2(1270)$ through the $\pi\pi\to\pi\pi$ partial-wave
scattering amplitude $t^0_2(s)$ over the $KK$ threshold, in the range
$\sqrt{s}_0\in [\, 1.15$~GeV$\, ,\, 1.40$~GeV$\,]$.
We safely assume this range as analytical,
since previous analysis find that the $\pi\pi$, $KK$ and $4\pi$ channels
provide more than $95\%$ of the $f_2$ branching ratio~\cite{PDG2012}.
In addition, the remaining observed decays ($\eta\eta$, $\eta\pi\pi$, $K^0K^-\pi^+$+c.c.,
$\gamma\gamma$, $e^+e-$) have branching ratios below $0.8\%$
and their thresholds are not in the $s_0$ range considered above.
Channels that could introduce  branch point singularities in that interval are
not observed.
Nonetheless, the inelasticity drops around the $f_2(1270)$ down to $\eta^0_2(s)\simeq 0.75$
and we cannot take $\eta^0_2(s)=1$ anymore in our analysis.
Therefore, we use the CFD parametrization  of the phase-shift $\delta^0_2(s)$ and inelasticity $\eta^0_2(s)$,
to compute the $\pi\pi$ tensor isoscalar partial wave $t^0_2(s)$~\cite{Martin:2011cn}.
Apart of this subtlety, our analysis of this channel proceeds exactly in the same way as
we did for the $f_0(500)$ and $\rho(770)$.
If the experimental uncertainties are dropped, the optimal PA center  for the $P^3_1(s,s_0)$
approximant is found to be $\sqrt{s_0}=1180$~MeV.
Once the statistical errors are incorporated through the MC described
previously (now generating also values for $\{\eta^{(n)}\}$ in the MC),
the total error turns minimal for $\sqrt{s}_0=1270$~MeV,
producing the pole position in the 3rd Riemann sheet (3RS) and coupling to two pions ($|Z^{(3)}=0.184\pm 0.019$~GeV$^2$) given in
Table~\ref{tab:f2-pole}, where we also add for comparison further $f_2(1270)$
pole determinations. 
It is particularly interesting to compare our determination with the results of~\cite{Dai:2014zta}, where the $f_2(1270)$ pole position was obtained from the process
$\gamma\gamma\to \pi\pi$, and the $\pi\pi$ CFD parameterizations of
\cite{GarciaMartin:2011jx} was also used as input. 
Despite the absence of errors in~\cite{Dai:2014zta}, 
the comparison between both results shows again a nice agreement.
The theoretical PA and the input uncertainties are found to be
of the same order of magnitude.

\begin{table}[!t]   
\caption{{\small    Collection of different predictions for the $f_2(1270)$ meson
resonance parameters in the 3RS.}}
\begin{center}
\begin{tabular}{ccc}
\hline
Reference& $\sqrt{s_{f_2}}$ (MeV) & $   g_{f_2\pi\pi}$  (GeV$^{-1}$) 
\\\hline
 \cite{Johnson:1968df} 
& $(1268\pm 6)- i(88\pm 7)$ 
&  --  \\
 \cite{Longacre:1986fh} 
&  $(1283^{+6}_{-5})-i(93^{+5}_{-1})$ 
&  -- \\
 \cite{Bertin:1997kh}  
&  $(1278\pm 5)-i(102\pm 10)$ 
&   --   \\
  \cite{Shchegelsky:2006et}  
&  $(1277\pm 6)-i(98\pm 8) $ 
&   --    \\
 \cite{Anisovich:2009zza} 
&  $(1270\pm 8)-i(97\pm 18)$  
&    --    \\
 \cite{Dai:2014zta}
& $1267-i108$ &--\\
\hline
\mbox{This Work} 
& $(1267.3\pm     1.7   )-i(95.0\pm 2.4)    $
&   $4.41\pm 0.23$     \\
\hline
\end{tabular}
\end{center}
\label{tab:f2-pole}
\end{table}

The next resonance in the $IJ=00$ channel is the $f_0(980)$.
Its determination is more cumbersome as it is placed close to the $KK$ threshold.
Both the phase-shift and the inelasticity vary very quickly and
an accurate determination of their first derivatives is rather complicated.
In addition, the $KK$ threshold puts a limit to the range of applicability of
Montessus' theorem in the $s$--variable.
This pathology  can be cured by working in the $k_K=\sqrt{s/4-m_K^2}$
variable or with a conformal mapping $\omega(s,s_0)$~\cite{Martin:2011cn,Masjuan:2013jha}.
Nevertheless, in spite of being a relatively narrow resonance, we find that our $PA$
sequences yield very unstable pole determinations:
using data from different energies in the range $\sqrt{s_0}= 900$--$1100$~MeV
gives place to displacements  in the position of orders of magnitude
and sometimes to different Riemann sheets --including 1RS--.
No conclusive  result was obtained from PA sequences
with different numbers of poles, even reducing the range
of analyzed data to $\sqrt{s_0}= 950$--$1010$~MeV.

Before concluding, we reassess the accuracy of our results by considering
extensions of Montessus' theorem. Beyond the $P^N_1(s,s_0)$ sequence, such
theorem also ensures convergence for the $P^N_2(s,s_0)$ even though only one
resonance pole would lie in the convergence disk provided that the second PA
pole lies outside of it, as expected. With four experimental derivatives, we
can go up to the $P^2_2(s,s_0)$, with a theoretical error defined as the
difference between the $P^1_2(s,s_0)$ and $P^2_2(s,s_0)$ pole
determinations. The results for $f_0(500), \rho(770)$, and $f_2(1270)$ are
collected in Table~\ref{tab:P22pred} and they are found to be in agreement
with the more conservative $P_1^N$ determinations given in
Tables~\ref{tab:sigma-pole}--\ref{tab:f2-pole}.

\begin{table}[!t]   
\caption{{\small    Collection of predictions using the $P^2_2(s,s_0)$ approximant.     }}
\begin{center}
\begin{tabular}{ccc}
\hline
 $P^2_2(s,s_0)$ & optimal $\sqrt{s_0}$ (MeV) & $\sqrt{s_P}$ (MeV)\\
\hline
$f_0(500)$ & $490$ & $(461\pm 13)-i(300\pm 11)$\\
$\rho(770)$ & $740$ &$(761.4\pm 0.8)-i(71.7\pm 0.7)$\\
$f_2(1270)$ &  $1240$ & $(1268.0\pm 1.7)-i(95.7\pm 1.8)$\\
\hline
\end{tabular}
\end{center}
\label{tab:P22pred}
\end{table}

\section{Conclusions}\label{sec:conclusions}
We have performed a safe and accurate determination of the lightest resonance pole parameters in the channels $IJ=00,11,02$, respectively,
$f_0(500)$, $\rho(770)$, and $f_2(1270)$, by using Pad\'e approximants to analytically
extend the CFD  $\pi\pi$-scattering  parameterizations of~\cite{Martin:2011cn} from
real energies into the complex plane. With such method, we extract the pole position with a level
of precision comparable to other approaches, keeping good control
of both the experimental uncertainties stemming form the GKPY input and the theoretical
uncertainties deriving from the Pad\'e approximant analytical extension. More precise experimental scattering
data, even for a very short energy range, could easily improve our determination of the resonance parameters:
in addition to a smaller statistical error, one could safely extract
a higher number of derivatives with appropriate precision and construct higher order PAs,
hence decreasing the theoretical error.

\section*{Acknowledgments:}
This work has been partially supported by
the Spanish Government and ERDF funds from the European Commission
[FPA2010-17747, FPA2013-44773-P,
"Centro de Excelencia Severo Ochoa" Programme under grant SEV-2012-0249,
Consolider-Ingenio CPAN CSD2007-00042],
the Comunidad de Madrid [HEPHACOS S2009/ESP-1473],
the MICINN-INFN fund AIC-D-2011-0818 and by the Deutsche Forschungsgemeinschaft DFG
through the Collaborative Research Center ``The Low-Energy Frontier of the Standard Model"
(SFB 1044) and "Subnuclear Structure of Maetter (SFB/TR 16)".


\begin{thebibliography}{22}%
\makeatletter
\providecommand \@ifxundefined [1]{%
 \@ifx{#1\undefined}
}%
\providecommand \@ifnum [1]{%
 \ifnum #1\expandafter \@firstoftwo
 \else \expandafter \@secondoftwo
 \fi
}%
\providecommand \@ifx [1]{%
 \ifx #1\expandafter \@firstoftwo
 \else \expandafter \@secondoftwo
 \fi
}%
\providecommand \natexlab [1]{#1}%
\providecommand \enquote  [1]{``#1''}%
\providecommand \bibnamefont  [1]{#1}%
\providecommand \bibfnamefont [1]{#1}%
\providecommand \citenamefont [1]{#1}%
\providecommand \href@noop [0]{\@secondoftwo}%
\providecommand \href [0]{\begingroup \@sanitize@url \@href}%
\providecommand \@href[1]{\@@startlink{#1}\@@href}%
\providecommand \@@href[1]{\endgroup#1\@@endlink}%
\providecommand \@sanitize@url [0]{\catcode `\\12\catcode `\$12\catcode
  `\&12\catcode `\#12\catcode `\^12\catcode `\_12\catcode `\%12\relax}%
\providecommand \@@startlink[1]{}%
\providecommand \@@endlink[0]{}%
\providecommand \url  [0]{\begingroup\@sanitize@url \@url }%
\providecommand \@url [1]{\endgroup\@href {#1}{\urlprefix }}%
\providecommand \urlprefix  [0]{URL }%
\providecommand \Eprint [0]{\href }%
\providecommand \doibase [0]{http://dx.doi.org/}%
\providecommand \selectlanguage [0]{\@gobble}%
\providecommand \bibinfo  [0]{\@secondoftwo}%
\providecommand \bibfield  [0]{\@secondoftwo}%
\providecommand \translation [1]{[#1]}%
\providecommand \BibitemOpen [0]{}%
\providecommand \bibitemStop [0]{}%
\providecommand \bibitemNoStop [0]{.\EOS\space}%
\providecommand \EOS [0]{\spacefactor3000\relax}%
\providecommand \BibitemShut  [1]{\csname bibitem#1\endcsname}%
\let\auto@bib@innerbib\@empty
\bibitem [{\citenamefont {G.A.Baker}\ and\ \citenamefont
  {Graves-Morris}(1996)}]{Baker}%
  \BibitemOpen
  \bibfield  {author} {\bibinfo {author} {\bibnamefont {G.A.Baker}}\ and\
  \bibinfo {author} {\bibfnamefont {P.}~\bibnamefont {Graves-Morris}},\
  }\href@noop {} {\emph {\bibinfo {title} {Pade Approximants}}},\ edited by\
  \bibinfo {editor} {\bibfnamefont {C.~U.}\ \bibnamefont {Press}}\ (\bibinfo
  {publisher} {Encyclopedia of Mathematics and its Applications},\ \bibinfo
  {year} {1996})\BibitemShut {NoStop}%
\bibitem [{\citenamefont {de~Montessus~de Ballore}(1902)}]{Montessus}%
  \BibitemOpen
  \bibfield  {author} {\bibinfo {author} {\bibfnamefont {R.}~\bibnamefont
  {de~Montessus~de Ballore}},\ }\href@noop {} {\bibfield  {journal} {\bibinfo
  {journal} {Bull. Soc. Math. France}\ }\textbf {\bibinfo {volume} {30}},\
  \bibinfo {pages} {28} (\bibinfo {year} {1902})}\BibitemShut {NoStop}%
\bibitem [{\citenamefont {Masjuan}\ and\ \citenamefont
  {Sanz-Cillero}(2013)}]{Masjuan:2013jha}%
  \BibitemOpen
  \bibfield  {author} {\bibinfo {author} {\bibfnamefont {P.}~\bibnamefont
  {Masjuan}}\ and\ \bibinfo {author} {\bibfnamefont {J.}~\bibnamefont
  {Sanz-Cillero}},\ }\href@noop {} {\bibfield  {journal} {\bibinfo  {journal}
  {Eur.Phys.J.}\ }\textbf {\bibinfo {volume} {C73}},\ \bibinfo {pages} {2594}
  (\bibinfo {year} {2013})},\ \Eprint {http://arxiv.org/abs/1306.6308}
  {arXiv:1306.6308 [hep-ph]} \BibitemShut {NoStop}%
\bibitem [{\citenamefont {Colangelo}\ \emph {et~al.}(2001)\citenamefont
  {Colangelo}, \citenamefont {Gasser},\ and\ \citenamefont
  {Leutwyler}}]{Colangelo:2001df}%
  \BibitemOpen
  \bibfield  {author} {\bibinfo {author} {\bibfnamefont {G.}~\bibnamefont
  {Colangelo}}, \bibinfo {author} {\bibfnamefont {J.}~\bibnamefont {Gasser}}, \
  and\ \bibinfo {author} {\bibfnamefont {H.}~\bibnamefont {Leutwyler}},\ }\href
  {\doibase 10.1016/S0550-3213(01)00147-X} {\bibfield  {journal} {\bibinfo
  {journal} {Nucl. Phys.}\ }\textbf {\bibinfo {volume} {B603}},\ \bibinfo
  {pages} {125} (\bibinfo {year} {2001})},\ \Eprint
  {http://arxiv.org/abs/hep-ph/0103088} {arXiv:hep-ph/0103088} \BibitemShut
  {NoStop}%
\bibitem [{\citenamefont {Buettiker}\ \emph {et~al.}(2004)\citenamefont
  {Buettiker}, \citenamefont {Descotes-Genon},\ and\ \citenamefont
  {Moussallam}}]{Moussallam:2004}%
  \BibitemOpen
  \bibfield  {author} {\bibinfo {author} {\bibfnamefont {P.}~\bibnamefont
  {Buettiker}}, \bibinfo {author} {\bibfnamefont {S.}~\bibnamefont
  {Descotes-Genon}}, \ and\ \bibinfo {author} {\bibfnamefont {B.}~\bibnamefont
  {Moussallam}},\ }\href {\doibase 10.1140/epjc/s2004-01591-1} {\bibfield
  {journal} {\bibinfo  {journal} {Eur.Phys.J.}\ }\textbf {\bibinfo {volume}
  {C33}},\ \bibinfo {pages} {409} (\bibinfo {year} {2004})},\ \Eprint
  {http://arxiv.org/abs/hep-ph/0310283} {arXiv:hep-ph/0310283 [hep-ph]}
  \BibitemShut {NoStop}%
\bibitem [{\citenamefont {Caprini}\ \emph {et~al.}(2006)\citenamefont
  {Caprini}, \citenamefont {Colangelo},\ and\ \citenamefont
  {Leutwyler}}]{Caprini:2005zr}%
  \BibitemOpen
  \bibfield  {author} {\bibinfo {author} {\bibfnamefont {I.}~\bibnamefont
  {Caprini}}, \bibinfo {author} {\bibfnamefont {G.}~\bibnamefont {Colangelo}},
  \ and\ \bibinfo {author} {\bibfnamefont {H.}~\bibnamefont {Leutwyler}},\
  }\href {\doibase 10.1103/PhysRevLett.96.132001} {\bibfield  {journal}
  {\bibinfo  {journal} {Phys. Rev. Lett.}\ }\textbf {\bibinfo {volume} {96}},\
  \bibinfo {pages} {132001} (\bibinfo {year} {2006})},\ \Eprint
  {http://arxiv.org/abs/hep-ph/0512364} {arXiv:hep-ph/0512364} \BibitemShut
  {NoStop}%
\bibitem [{\citenamefont {Descotes-Genon}\ and\ \citenamefont
  {Moussallam}(2006)}]{DescotesGenon:2006uk}%
  \BibitemOpen
  \bibfield  {author} {\bibinfo {author} {\bibfnamefont {S.}~\bibnamefont
  {Descotes-Genon}}\ and\ \bibinfo {author} {\bibfnamefont {B.}~\bibnamefont
  {Moussallam}},\ }\href {\doibase 10.1140/epjc/s10052-006-0036-2} {\bibfield
  {journal} {\bibinfo  {journal} {Eur.Phys.J.}\ }\textbf {\bibinfo {volume}
  {C48}},\ \bibinfo {pages} {553} (\bibinfo {year} {2006})},\ \Eprint
  {http://arxiv.org/abs/hep-ph/0607133} {arXiv:hep-ph/0607133 [hep-ph]}
  \BibitemShut {NoStop}%
\bibitem [{\citenamefont {Garcia-Martin}\ \emph
  {et~al.}(2011{\natexlab{a}})\citenamefont {Garcia-Martin}, \citenamefont
  {Kaminski}, \citenamefont {Pelaez}, \citenamefont {Ruiz~de Elvira},\ and\
  \citenamefont {Yndurain}}]{Martin:2011cn}%
  \BibitemOpen
  \bibfield  {author} {\bibinfo {author} {\bibfnamefont {R.}~\bibnamefont
  {Garcia-Martin}}, \bibinfo {author} {\bibfnamefont {R.}~\bibnamefont
  {Kaminski}}, \bibinfo {author} {\bibfnamefont {J.}~\bibnamefont {Pelaez}},
  \bibinfo {author} {\bibfnamefont {J.}~\bibnamefont {Ruiz~de Elvira}}, \ and\
  \bibinfo {author} {\bibfnamefont {F.}~\bibnamefont {Yndurain}},\ }\href
  {\doibase 10.1103/PhysRevD.83.074004} {\bibfield  {journal} {\bibinfo
  {journal} {Phys.Rev.}\ }\textbf {\bibinfo {volume} {D83}},\ \bibinfo {pages}
  {074004} (\bibinfo {year} {2011}{\natexlab{a}})},\ \Eprint
  {http://arxiv.org/abs/1102.2183} {arXiv:1102.2183 [hep-ph]} \BibitemShut
  {NoStop}%
\bibitem [{\citenamefont {Garcia-Martin}\ \emph
  {et~al.}(2011{\natexlab{b}})\citenamefont {Garcia-Martin}, \citenamefont
  {Kaminski}, \citenamefont {Pelaez},\ and\ \citenamefont {Ruiz~de
  Elvira}}]{GarciaMartin:2011jx}%
  \BibitemOpen
  \bibfield  {author} {\bibinfo {author} {\bibfnamefont {R.}~\bibnamefont
  {Garcia-Martin}}, \bibinfo {author} {\bibfnamefont {R.}~\bibnamefont
  {Kaminski}}, \bibinfo {author} {\bibfnamefont {J.}~\bibnamefont {Pelaez}}, \
  and\ \bibinfo {author} {\bibfnamefont {J.}~\bibnamefont {Ruiz~de Elvira}},\
  }\href {\doibase 10.1103/PhysRevLett.107.072001} {\bibfield  {journal}
  {\bibinfo  {journal} {Phys.Rev.Lett.}\ }\textbf {\bibinfo {volume} {107}},\
  \bibinfo {pages} {072001} (\bibinfo {year} {2011}{\natexlab{b}})},\ \Eprint
  {http://arxiv.org/abs/1107.1635} {arXiv:1107.1635 [hep-ph]} \BibitemShut
  {NoStop}%
\bibitem [{\citenamefont {Ditsche}\ \emph {et~al.}(2012)\citenamefont
  {Ditsche}, \citenamefont {Hoferichter}, \citenamefont {Kubis},\ and\
  \citenamefont {Meissner}}]{Ditsche:2012fv}%
  \BibitemOpen
  \bibfield  {author} {\bibinfo {author} {\bibfnamefont {C.}~\bibnamefont
  {Ditsche}}, \bibinfo {author} {\bibfnamefont {M.}~\bibnamefont
  {Hoferichter}}, \bibinfo {author} {\bibfnamefont {B.}~\bibnamefont {Kubis}},
  \ and\ \bibinfo {author} {\bibfnamefont {U.-G.}\ \bibnamefont {Meissner}},\
  }\href {\doibase 10.1007/JHEP06(2012)043} {\bibfield  {journal} {\bibinfo
  {journal} {JHEP}\ }\textbf {\bibinfo {volume} {1206}},\ \bibinfo {pages}
  {043} (\bibinfo {year} {2012})},\ \Eprint {http://arxiv.org/abs/1203.4758}
  {arXiv:1203.4758 [hep-ph]} \BibitemShut {NoStop}%
\bibitem [{\citenamefont {Batley}\ \emph {et~al.}(2010)\citenamefont {Batley}
  \emph {et~al.}}]{Batley:2010zza}%
  \BibitemOpen
  \bibfield  {author} {\bibinfo {author} {\bibfnamefont {J.}~\bibnamefont
  {Batley}} \emph {et~al.} (\bibinfo {collaboration} {NA48-2 Collaboration}),\
  }\href {\doibase 10.1140/epjc/s10052-010-1480-6} {\bibfield  {journal}
  {\bibinfo  {journal} {Eur.Phys.J.}\ }\textbf {\bibinfo {volume} {C70}},\
  \bibinfo {pages} {635} (\bibinfo {year} {2010})}\BibitemShut {NoStop}%
\bibitem [{\citenamefont {Zhou}\ \emph {et~al.}(2005)\citenamefont {Zhou},
  \citenamefont {Qin}, \citenamefont {Zhang}, \citenamefont {Xiao},
  \citenamefont {Zheng} \emph {et~al.}}]{rho-Zheng}%
  \BibitemOpen
  \bibfield  {author} {\bibinfo {author} {\bibfnamefont {Z.}~\bibnamefont
  {Zhou}}, \bibinfo {author} {\bibfnamefont {G.}~\bibnamefont {Qin}}, \bibinfo
  {author} {\bibfnamefont {P.}~\bibnamefont {Zhang}}, \bibinfo {author}
  {\bibfnamefont {Z.}~\bibnamefont {Xiao}}, \bibinfo {author} {\bibfnamefont
  {H.}~\bibnamefont {Zheng}},  \emph {et~al.},\ }\href {\doibase
  10.1088/1126-6708/2005/02/043} {\bibfield  {journal} {\bibinfo  {journal}
  {JHEP}\ }\textbf {\bibinfo {volume} {0502}},\ \bibinfo {pages} {043}
  (\bibinfo {year} {2005})},\ \Eprint {http://arxiv.org/abs/hep-ph/0406271}
  {arXiv:hep-ph/0406271 [hep-ph]} \BibitemShut {NoStop}%
\bibitem [{\citenamefont {Moussallam}(2011)}]{Moussallam:2011zg}%
  \BibitemOpen
  \bibfield  {author} {\bibinfo {author} {\bibfnamefont {B.}~\bibnamefont
  {Moussallam}},\ }\href {\doibase 10.1140/epjc/s10052-011-1814-z} {\bibfield
  {journal} {\bibinfo  {journal} {Eur.Phys.J.}\ }\textbf {\bibinfo {volume}
  {C71}},\ \bibinfo {pages} {1814} (\bibinfo {year} {2011})},\ \Eprint
  {http://arxiv.org/abs/1110.6074} {arXiv:1110.6074 [hep-ph]} \BibitemShut
  {NoStop}%
\bibitem [{\citenamefont {Ananthanarayan}\ \emph {et~al.}(2001)\citenamefont
  {Ananthanarayan}, \citenamefont {Colangelo}, \citenamefont {Gasser},\ and\
  \citenamefont {Leutwyler}}]{Roy}%
  \BibitemOpen
  \bibfield  {author} {\bibinfo {author} {\bibfnamefont {B.}~\bibnamefont
  {Ananthanarayan}}, \bibinfo {author} {\bibfnamefont {G.}~\bibnamefont
  {Colangelo}}, \bibinfo {author} {\bibfnamefont {J.}~\bibnamefont {Gasser}}, \
  and\ \bibinfo {author} {\bibfnamefont {H.}~\bibnamefont {Leutwyler}},\
  }\href@noop {} {\bibfield  {journal} {\bibinfo  {journal} {Phys. Rept.}\
  }\textbf {\bibinfo {volume} {353}},\ \bibinfo {pages} {207} (\bibinfo {year}
  {2001})},\ \Eprint {http://arxiv.org/abs/hep-ph/0005297} {hep-ph/0005297}
  \BibitemShut {NoStop}%
\bibitem [{\citenamefont {Sanz-Cillero}\ and\ \citenamefont
  {Pich}(2003)}]{Cillero-VFF}%
  \BibitemOpen
  \bibfield  {author} {\bibinfo {author} {\bibfnamefont {J.}~\bibnamefont
  {Sanz-Cillero}}\ and\ \bibinfo {author} {\bibfnamefont {A.}~\bibnamefont
  {Pich}},\ }\href {\doibase 10.1140/epjc/s2002-01128-8} {\bibfield  {journal}
  {\bibinfo  {journal} {Eur.Phys.J.}\ }\textbf {\bibinfo {volume} {C27}},\
  \bibinfo {pages} {587} (\bibinfo {year} {2003})},\ \Eprint
  {http://arxiv.org/abs/hep-ph/0208199} {arXiv:hep-ph/0208199 [hep-ph]}
  \BibitemShut {NoStop}%
\bibitem [{\citenamefont {Beringer}\ \emph {et~al.}(2012)\citenamefont
  {Beringer} \emph {et~al.}}]{PDG2012}%
  \BibitemOpen
  \bibfield  {author} {\bibinfo {author} {\bibfnamefont {J.}~\bibnamefont
  {Beringer}} \emph {et~al.} (\bibinfo {collaboration} {Particle Data Group}),\
  }\href {\doibase 10.1103/PhysRevD.86.010001} {\bibfield  {journal} {\bibinfo
  {journal} {Phys.Rev.}\ }\textbf {\bibinfo {volume} {D86}},\ \bibinfo {pages}
  {010001} (\bibinfo {year} {2012})}\BibitemShut {NoStop}%
\bibitem [{\citenamefont {Dai}\ and\ \citenamefont
  {Pennington}(2014)}]{Dai:2014zta}%
  \BibitemOpen
  \bibfield  {author} {\bibinfo {author} {\bibfnamefont {L.-Y.}\ \bibnamefont
  {Dai}}\ and\ \bibinfo {author} {\bibfnamefont {M.~R.}\ \bibnamefont
  {Pennington}},\ }\href {\doibase 10.1103/PhysRevD.90.036004} {\bibfield
  {journal} {\bibinfo  {journal} {Phys.Rev.}\ }\textbf {\bibinfo {volume}
  {D90}},\ \bibinfo {pages} {036004} (\bibinfo {year} {2014})},\ \Eprint
  {http://arxiv.org/abs/1404.7524} {arXiv:1404.7524 [hep-ph]} \BibitemShut
  {NoStop}%
\bibitem [{\citenamefont {Johnson}\ \emph {et~al.}(1968)\citenamefont {Johnson}
  \emph {et~al.}}]{Johnson:1968df}%
  \BibitemOpen
  \bibfield  {author} {\bibinfo {author} {\bibfnamefont {P.}~\bibnamefont
  {Johnson}} \emph {et~al.} (\bibinfo {collaboration} {Notre Dame-Purdue-SLAC
  Collaboration}),\ }\href {\doibase 10.1103/PhysRev.176.1651} {\bibfield
  {journal} {\bibinfo  {journal} {Phys.Rev.}\ }\textbf {\bibinfo {volume}
  {176}},\ \bibinfo {pages} {1651} (\bibinfo {year} {1968})}\BibitemShut
  {NoStop}%
\bibitem [{\citenamefont {Longacre}\ \emph {et~al.}(1986)\citenamefont
  {Longacre}, \citenamefont {Etkin}, \citenamefont {Foley}, \citenamefont
  {Love}, \citenamefont {Morris} \emph {et~al.}}]{Longacre:1986fh}%
  \BibitemOpen
  \bibfield  {author} {\bibinfo {author} {\bibfnamefont {R.}~\bibnamefont
  {Longacre}}, \bibinfo {author} {\bibfnamefont {A.}~\bibnamefont {Etkin}},
  \bibinfo {author} {\bibfnamefont {K.}~\bibnamefont {Foley}}, \bibinfo
  {author} {\bibfnamefont {W.}~\bibnamefont {Love}}, \bibinfo {author}
  {\bibfnamefont {T.}~\bibnamefont {Morris}},  \emph {et~al.},\ }\href
  {\doibase 10.1016/0370-2693(86)91061-0} {\bibfield  {journal} {\bibinfo
  {journal} {Phys.Lett.}\ }\textbf {\bibinfo {volume} {B177}},\ \bibinfo
  {pages} {223} (\bibinfo {year} {1986})}\BibitemShut {NoStop}%
\bibitem [{\citenamefont {Bertin}\ \emph {et~al.}(1997)\citenamefont {Bertin}
  \emph {et~al.}}]{Bertin:1997kh}%
  \BibitemOpen
  \bibfield  {author} {\bibinfo {author} {\bibfnamefont {A.}~\bibnamefont
  {Bertin}} \emph {et~al.} (\bibinfo {collaboration} {OBELIX Collaboration}),\
  }\href {\doibase 10.1016/S0370-2693(97)00791-0} {\bibfield  {journal}
  {\bibinfo  {journal} {Phys.Lett.}\ }\textbf {\bibinfo {volume} {B408}},\
  \bibinfo {pages} {476} (\bibinfo {year} {1997})}\BibitemShut {NoStop}%
\bibitem [{\citenamefont {Shchegelsky}\ \emph {et~al.}(2006)\citenamefont
  {Shchegelsky}, \citenamefont {Sarantsev}, \citenamefont {Nikonov},\ and\
  \citenamefont {Anisovich}}]{Shchegelsky:2006et}%
  \BibitemOpen
  \bibfield  {author} {\bibinfo {author} {\bibfnamefont {V.}~\bibnamefont
  {Shchegelsky}}, \bibinfo {author} {\bibfnamefont {A.}~\bibnamefont
  {Sarantsev}}, \bibinfo {author} {\bibfnamefont {V.}~\bibnamefont {Nikonov}},
  \ and\ \bibinfo {author} {\bibfnamefont {A.}~\bibnamefont {Anisovich}},\
  }\href {\doibase 10.1140/epja/i2005-10264-2} {\bibfield  {journal} {\bibinfo
  {journal} {Eur.Phys.J.}\ }\textbf {\bibinfo {volume} {A27}},\ \bibinfo
  {pages} {207} (\bibinfo {year} {2006})}\BibitemShut {NoStop}%
\bibitem [{\citenamefont {Anisovich}\ and\ \citenamefont
  {Sarantsev}(2009)}]{Anisovich:2009zza}%
  \BibitemOpen
  \bibfield  {author} {\bibinfo {author} {\bibfnamefont {V.}~\bibnamefont
  {Anisovich}}\ and\ \bibinfo {author} {\bibfnamefont {A.}~\bibnamefont
  {Sarantsev}},\ }\href {\doibase 10.1142/S0217751X09043286} {\bibfield
  {journal} {\bibinfo  {journal} {Int.J.Mod.Phys.}\ }\textbf {\bibinfo {volume}
  {A24}},\ \bibinfo {pages} {2481} (\bibinfo {year} {2009})}\BibitemShut
  {NoStop}%
\end{thebibliography}
\end{document}